\documentclass{ws-mplb}

\newcommand{\bra}[1]{\ensuremath{\langle #1 |}}
\newcommand{\ket}[1]{\ensuremath{| #1 \rangle}}
\newcommand{\expect}[1]{\langle \rangle}

\usepackage{amsmath}
\usepackage{amssymb}

\begin{document}

\markboth{S. Capponi, F. Alet, and M. Mambrini}{ENTANGLEMENT OF QUANTUM SPIN SYSTEMS AND VALENCE BONDS}

%
\catchline{}{}{}{}{}
%

\title{ENTANGLEMENT OF QUANTUM SPIN SYSTEMS: A VALENCE-BOND APPROACH}

\author{\footnotesize SYLVAIN CAPPONI, FABIEN ALET, and MATTHIEU MAMBRINI}
\address{
Laboratoire de Physique Th\'eorique, Universit\'e de Toulouse \& CNRS, UPS (IRSAMC), F-31062 Toulouse, France\\capponi,alet,mambrini@irsamc.ups-tlse.fr}

\maketitle

\begin{history}
\received{(Day Month Year)}
\revised{(Day Month Year)}
\end{history}

\begin{abstract}
In order to quantify entanglement between two parts of a quantum system, one of the most used estimator is the Von Neumann entropy. Unfortunately, computing this quantity for large interacting quantum spin systems remains an open issue. Faced with this difficulty, other estimators have been proposed to measure entanglement efficiently, mostly by using simulations in the valence-bond basis. We review the different proposals and try to clarify the connections between their geometric definitions and proper observables. We illustrate this analysis with new results of entanglement properties of spin 1 chains. 
\end{abstract}


\section{Introduction}

Entanglement is a fundamental notion of quantum mechanics, that has over the recent years gained popularity as a way to provide new insights in the quantum many-body problem. From
the condensed matter point of view, one of the most interesting
promises of the study of entanglement properties is the possibility to
automatically detect the nature of quantum phases and of quantum phase
transitions.  In this approach, there is no need to provide {\it a
  priori} physical information or input, such as the specification of
an order parameter. The detection can occur through the study of the
scaling (with system size) of various entanglement estimators. For
instance, the scaling of the von Neumann entanglement entropy for
one-dimensional systems is different for critical and gapped systems -
allowing their distinction. For a recent review of various properties
of entanglement entropy in condenser matter, see Ref.~\protect\refcite{review}.

Unfortunately, computing von Neumann entropy for large interacting systems remains a highly difficult task, except in one dimension where it can easily be accessed with Density Matrix Renormalization Group (DMRG) algorithm. 
As a consequence, several other estimators have been proposed recently, that can be computed efficiently with large-scale quantum Monte Carlo (QMC) techniques. Most of these proposals use simulations in the valence-bond (VB) basis. 
In section \ref{VB.part}, we define useful properties of the VB basis. 
In section \ref{EE.part}, we review the different estimators for entanglement entropy and clarify their definitions, thus allowing to compute some of them with other techniques. 
Finally, in section \ref{num.part}, we present results for frustrated spin-1/2 and spin-1 chains.

\section{Valence-Bond Basis}\label{VB.part}


Let us first consider the  antiferromagnetic (AF)  
spin-$1/2$ Heisenberg Hamiltonian
\begin{equation*}
H=\sum_{ij} J_{ij} {\bf S}_i\cdot {\bf S}_j,
\label{eq:H}
\end{equation*}
which conserves the total spin $S_{\rm T}$ of the system. AF interactions $J_{ij}>0$
favor a singlet $S_{\rm T}=0$ ground-state.

It is well-known that any singlet state can be expressed in
the VB basis, where spins couple pairwise in singlets $(\ket{\!\! \uparrow
  \downarrow}-\ket{\!\! \downarrow \uparrow})/\sqrt{2}$. The VB basis is
overcomplete (there are more VB coverings than the total number of
singlets). Another basis is the {\it bipartite} VB basis\cite{Beach06},
where the system is decomposed into two sets  such that two spins
forming a singlet necessarily belong to different sets. This basis is smaller, but
still overcomplete. 
Choosing a bipartition of the $N$-sites lattice into two equal sized subsets $\cal A$ and $\cal B$ (for instance different sublattices for a bipartite lattice), then the bipartite VB subspace is generated by all the bipartite VB states
\begin{equation}
\label{eq:bip_rvb}
\vert \varphi_{\cal D} \rangle = \bigotimes_{\substack{(i,j) \in {\cal D} \\ i \in {\cal A}, j \in {\cal B}}} [i,j],
\end{equation}
where $[i,j]$ is a SU(2) dimer state and ${\cal D}$ is a dimer covering of the system.

A crucial property of the VB basis is its non-orthogonality. Indeed, for any pair of VB states $\vert \varphi_{\cal D} \rangle$ and $\vert \varphi_{{\cal D}'}\rangle$, the overlap is given by
\begin{equation}
\label{eq:nonorth}
\langle \varphi_{{\cal D}'} \vert \varphi_{\cal D} \rangle = \varepsilon \left ( {\cal D}, {\cal D}'\right ) 2^{n_l \left ( {\cal D}, {\cal D}'\right ) - N/2},
\end{equation}
where $n_l \left ( {\cal D}, {\cal D}'\right )$ is the number of loops of the overlap diagram obtained by superimposing the two dimer coverings ${\cal D}$ and ${\cal D}'$ and $\varepsilon \left ( {\cal D}, {\cal D}'\right )$ a sign directly related to the antisymmetric structure of the two-site singlet. Since all sites of the overlap diagram are involved in exactly one dimer from ${\cal D}$ and one dimer from ${\cal D}'$ only even sized closed loops are possible. Each loop produces a multiplicative factor $2$ to Eq.~(\ref{eq:nonorth}) since only the two sequences $\uparrow, \downarrow,\ldots,\downarrow$ and $\downarrow,\uparrow\ldots,\uparrow$ along the loop lead to a non-vanishing contribution. For arbitrary range VB states, the sign $\varepsilon \left ( {\cal D}, {\cal D}'\right )$ cannot be fixed to a constant independent of the couple $\left ( {\cal D}, {\cal D}' \right )$ by a generic convention. However, considering only bipartite VB states, a convenient choice consists in orienting the dimers from sublattice $\cal A$ to $\cal B$. In this case, it can be readilly checked that $\varepsilon \left ( {\cal D}, {\cal D}'\right )=+1$ hence making $\langle \varphi_{{\cal D}'} \vert \varphi_{\cal D} \rangle$ an unsigned quantity which turns out to be crucial for Monte-Carlo applications. 

Despite being an overcomplete and non-orthogonal basis, Sandvik has shown how to make large-scale QMC simulations in the VB basis\cite{Sandvik}, thus allowing to sample ground-state of non-frustrated spin models. The basic idea consists in applying the hamiltonian a large number of times on a trial state, in order to project out onto the ground-state of the system. Moreover, recently Sandvik 
and Evertz have improved this algorithm by implementing non-local moves.\cite{SandvikEvertz}

\section{Measuring Entanglement}\label{EE.part}

\subsection{Generalities}

Over the past few years, various measures of entanglement have been used to investigate quantum phase
transitions and states of matter\cite{entanglement,Vidal,Calabrese04,Refael04}.
The most famous estimator is the so-called von Neumann entanglement entropy ($S^\mathrm{vN}$, see below for its definition) and in our context, one wants to compute how this entanglement entropy (EE) scales with the block size. 
While many interesting properties of EE have been derived exactly for
integrable models or systems with exactly known ground-states (GS), the
calculation of EE for general interacting quantum systems is an exacting
task. On the practical side, even numerical simulations are difficult. Exact diagonalization (ED) is limited to
small systems and cannot precisely verify scaling properties and EE is not accessible to QMC methods. In 1d, EE can be calculated within the DMRG method\cite{White92} and crucial information on the criticality can be extracted using conformal field theory tools.\cite{Calabrese04} Unfortunately, DMRG is not available in larger dimension. Therefore, several other estimators have been proposed to measure entanglement using QMC techniques in the VB basis. 

From the experimental point of view, if one considers small blocks (1 or 2 sites for instance), then the reduced density matrix (and thus EE) can be expressed in terms of all correlations on the same block: local magnetization, nearest-neighbor correlations \ldots so that in principle, local measurements can give access to these entropies. But, since we are mostly interested in large blocks to get insights on the nature of the quantum state, it seems impossible to be able to measure EE in this limit. 
Despite this negative result, Klich and Levitov have shown that, for free fermions, EE can be obtained from the cumulants of the number of transmitted charges distribution.~\cite{Klich} However, this result may not hold in the general interacting case.

\subsection{Different kinds of entropy}

Without trying to be exhaustive, let us discuss a few different measures of entanglement that are of interest for quantum spin systems, such as described by Eq.~(\ref{eq:H}).

\subsubsection{Von Neumann entanglement entropy}

The von Neumann EE ($S^\mathrm{vN}$) quantifies the bipartite entanglement
between two parts of a quantum system\cite{Bennett96}. For a quantum state $\ket{\Psi}$, EE 
between a part $\Omega$ and the rest of a system is $\displaystyle S_\Omega=-\mathrm{Tr} \rho_\Omega \ln \rho_\Omega $
where $\rho_\Omega=\mathrm{Tr}_{\bar{\Omega}}\ket{\Psi}\bra{\Psi}$ is the reduced density matrix of $\Omega$
obtained by tracing out the rest of the system $\bar{\Omega}$.
An important property of EE is that $S_\Omega=S_{\bar{\Omega}}$ and it appears
  naturally that $S_\Omega$ is only related to the common property of $\Omega$ and
  $\bar{\Omega}$, their boundary: general arguments indeed indicate that
  $S_\Omega$ typically scales with the size of the boundary (so called {\it area
    law}).\cite{area} 
For critical systems however, logarithmic corrections can be present. If the
critical system is conformal invariant, the amplitude of the logarithmic
corrections is related to the central charge of the corresponding
conformal field theory (CFT). This has been shown in one dimension
(1d)\cite{Calabrese04,Vidal} and for some CFT in two
dimensions (2d)\cite{Fradkin06}, in which case the
coefficient also depends on the geometry of $\Omega$. The scaling of the EE with the
size of $\Omega$ therefore contains precious informations on the state of the
physical system and can be used {\it e.g.} to detect criticality. This has
been successfully demonstrated for instance in quantum spin systems.\cite{Vidal,Calabrese04,Jin04,Laflo06}

\subsubsection{R\'enyi entropy}

In quantum information theory, the R\'enyi entropy is a generalization of $S^\mathrm{vN}$ for any integer $n>1$:
\begin{equation}\label{renyi.eq}
S_n = \frac{1}{1-n} \ln \mathrm{Tr} \rho_\Omega^n.
\end{equation}
In fact, $S^\mathrm{vN}$ can be viewed as a limit of $S_n$ when $n\rightarrow 1$. 

It turns out that R\'enyi entropy behaves quite similarly as $S^\mathrm{vN}$. In particular, for critical 1d systems, it has also logarithmic correction: for periodic boundary conditions (PBC) and a block of $x$ sites, one has\cite{Calabrese04,Vidal}
\begin{equation}
S_n = \frac{c}{6}(1+\frac{1}{n}) \ln x +K
\end{equation}
(with K a constant), thus allowing to extract the central charge $c$. 

Recently, a crucial point has been realized: it is possible to compute $S_n$ for $n>1$ with QMC in the VB basis\cite{Hastings}.
Despite a sampling problem that challenges the obtention of R\'enyi entropies for all blocks in the same simulation, the authors of Ref.~\protect\refcite{Hastings} have provided compelling evidence in favor of a strict area law for $S_2$ in the 2d Heisenberg case.

\subsubsection{Valence Bond entanglement entropy}
\label{sec:VBEE}
For a given bipartition of sites and bipartite VB state $\ket{\Phi}$, let us consider a subsystem
$\Omega$. One can define the Valence Bond EE ($S^{\rm VB}$) of this state as:
\begin{equation*}
S^{\rm VB}_\Omega(\ket{\Phi})=\ln(2) . \, n^c_\Omega(\ket{\Phi})
\end{equation*}
where $n^c_\Omega(\ket{\Phi})$ is {\it the number of singlets that cross the boundary of $\Omega$}.\cite{Alet,Chhajlany} The constant $\ln(2)$ is used to match the EE for a single site. 
  In general, the GS of the Hamiltonian $H$ is {\it not} a
  single VB state. For a linear combination $\ket{\Psi}=\sum_i a_i
\ket{\Phi_i}$ with $\ket{\Phi_i}$ bipartite VB states, we have defined 
$S^{\rm VB}_\Omega(\ket{\Psi})=\sum_i a_i
S^{\rm VB}_\Omega(\ket{\Phi_i})/\sum_i a_i$, which can thus be computed from the sampling of the  GS of $H$ by using QMC\cite{Sandvik,SandvikEvertz}. 

It turns out that $S^{\rm VB}$ has a similar behaviour as $S^{\rm vN}$ for gapless and gapped 1d systems (with respectively existence and absence of log corrections). In 2d, $S^{\rm VB}$ satisfies a pure area law for gapped ground-state but has multiplicative log corrections in the gapless N\'eel phase\cite{Alet,Chhajlany} that seems to differ from $S^\mathrm{vN}$ which has been claimed to satisfy a strict area law, according to DMRG simulations using  ladders with many legs.\cite{Kallin}

Recently, Lin and Sandvik have also proposed to measure a different VB EE by computing the average of the number of crossing links in the GS: $S^{\rm VB, 2}=\langle n^c_\Omega \rangle$. This quantity scales as $S^{\rm VB}$, but with smaller subleading scaling corrections for a spin chain.\cite{Sandvik2010}

\subsubsection{Loop EE}

For a single VB configuration, its loop graph is identical to the configuration itself. Therefore, EE is identically equal either to the number of bonds crossing the boundary, or to the number of loops crossing it. Using this idea, Lin and Sandvik have proposed to extend the definition to an arbitrary state by computing the average number of loops that cross a boundary between two blocks.\cite{Sandvik2010}

It turns out that in 1d, the scaling is logarithmic with the size of the block with a prefactor $\ln(2)/3$ for uniform chains with PBC, while in 2d, $S^{\rm loop}$ satisfies a strict area law for the N\'eel state.\cite{Sandvik2010}

\subsubsection{Other proposals}
If one considers a conserved quantity (like the total number of particles in the canonical ensemble, or the total magnetization for a spin system), then it could be useful to investigate the fluctuations of this quantity on a finite part of the system.\cite{LeHur}
However, contrary to EE, one needs to make a particular choice that may depend on the physics so that this approach looks less systematic.

Recently, it has also been proposed to look not only at the entropy, but at the whole distribution of eigenvalues of the reduced density matrix of a block, namely the entanglement spectrum\cite{Haldane2008}. Indeed, the entanglement spectrum of a block $\Omega$ appears to be related to the physical spectrum of the same Hamiltonian if the system was restricted to this block. Because of that, it can be used to investigate edge physics of quantum Hall effect even if the total system has no edges\cite{Regnault2009}. Concerning magnetic systems, this strategy has been pursued on spin-1/2 ladders by measuring the entanglement spectrum of one leg, that shows 1d-like features\cite{Poilblanc2010}. Unfortunately, because the full spectrum is needed, only ED techniques can be used, which does not allow to reach large systems. 

\subsection{Is it well defined ?}

While $S^{\rm vN}$ or Renyi entropies are clearly good quantum observables, $S^{\rm VB}$ and loop entropies are defined in terms of geometric objects (number of bonds or loops that cross a given boundary), but because of the overcompleteness of the VB basis, it is not granted that one would get the same results if the wavefunction was written down in a different (but equivalent) way. 

\subsubsection{VB entropy}
\label{sec:VBdef}
At first glance, $S^{\rm VB}$ seems ill-defined as the bipartite VB basis is
overcomplete. $\ket{\Psi}$ can indeed be rewritten as as a different linear combination
of other states for which $S^{\rm VB}$ could be different. However, $S^{\rm
  VB}$ turns out to be {\it conserved} in any linear combination between the
bipartite VB states. It can be indeed shown\cite{Mambrini} that for any linear relation $\sum_i c_i
\ket{\Phi_i}=0$ between the bipartite VB states $\ket{\Phi_i}$, we have $\sum_i c_i S^{\rm VB}_\Omega(\ket{\Phi_i})=0$ for all
$\Omega$.
Another issue is the choice of the bipartite
basis: in the general case, $S^{\rm VB}$ {\it does} depend on the precise choice of
bipartition. However, if $\ket{\Psi}$ satisfies a Marshall
sign criterion\cite{Marshall}, a genuine bipartition where all $a_i>0$ (in the notations of Sec.~\ref{sec:VBEE})
exists and should be taken. Not all singlet states
satisfy this, but this is the case for {\it e.g.} the
GS of Eq.~(\ref{eq:H}) on bipartite lattices.\cite{Marshall} 

An alternative definition of the occupation number, which moreover is also valid for any spin $S$, has been given in Ref.~\protect\refcite{Alet2010}. Let us first define a spin-$S$ dimer as the two-sites singlet state:
\begin{equation}
\label{eq:dimerS}
[i,j]_S = \frac{1}{\sqrt{2S+1}} \sum_{s_z=-S}^{+S} (-1)^{S-s_z} \vert -s_z,+s_z\rangle
\end{equation}
We also define the reference state $\vert R_S \rangle = \vert -S , +S , \ldots , -S , +S \rangle$
where the state is written in the $\otimes_i \hat{S}_z$ eigenstates basis and ordered such as ${\cal A}$ and ${\cal B}$ sites appear in alternating order. In particular $\vert R_{1/2} \rangle$ is nothing but the N\'eel state. As already noticed\cite{Sandvik} for $S=1/2$, the reference state has an equal overlap with all bipartite VB states: $\langle R_S \vert \varphi_{\cal D}\rangle$ does not depend on the bipartite dimer covering $\cal D$. This property is established by a direct evaluation leading to $\langle R_S \vert \varphi_{\cal D}\rangle= 1/(2S+1)^{N/4}$, with $N$ the total number of spins in the system.

After some algebra\cite{Alet2010}, one obtains that the ``average'' number of VBs between site $i$ and $j$ is given by:
\begin{equation}
\label{eq:projection_mixed_normalized}
n_{(i,j)}(\vert \Psi \rangle)= - \frac{1}{2S} \frac{\langle R_S \vert \hat{S}^{+}_i \hat{S}^{-}_j \vert \Psi
 \rangle}{\langle R_S  \vert \Psi \rangle}.
\end{equation}
The total number of VBs crossing the boundary of a given bipartition $\Omega$, and therefore $S^{\rm VB}$, are then easily  obtained by summing over all $i$ in $\Omega$ and $j$ in $\bar{\Omega}$. Note that this formulation is {\em explicitly independent} of the
 linear combination chosen to expand $\vert \Psi \rangle$ on the overcomplete bipartite VB manifold as it only involves projections of $\vert \Psi \rangle$. Moreover, this definition of $S^\mathrm{VB}$ as a projection is VB basis-free and therefore allows numerical computations  
outside the VB QMC scheme such as with ED and DMRG (see Sec.~\ref{num.part}).
 
 In summary,  $S^{\rm VB}$ is a perfectly well-defined quantity for any singlet wave function, although it does depend on the choice of bipartition. 

\subsubsection{Loop entropy}

The previous discussion emphasizes the fact that any quantity designed for measuring entanglement should be checked first to be consistently defined with respect to quantum mechanics. In particular, for a given singlet state, it should not be tied to a particular VB representation among all the possible choices made possible by the massive overcompletness of the (bipartite) VB basis. As obvious as such a requirement may seem, it is worth mentioning that since the various proposals have been introduced in the context of QMC\cite{Alet,Chhajlany,Sandvik2010}, they are constructed as {\em statistical averages} which do not necessarily make them properly defined as {\em quantum mechanical averages} (expectation value of an hermitian operator) or {\em projections} (like $S^{\rm VB}$).

\begin{figure}[th]
\centerline{\psfig{file=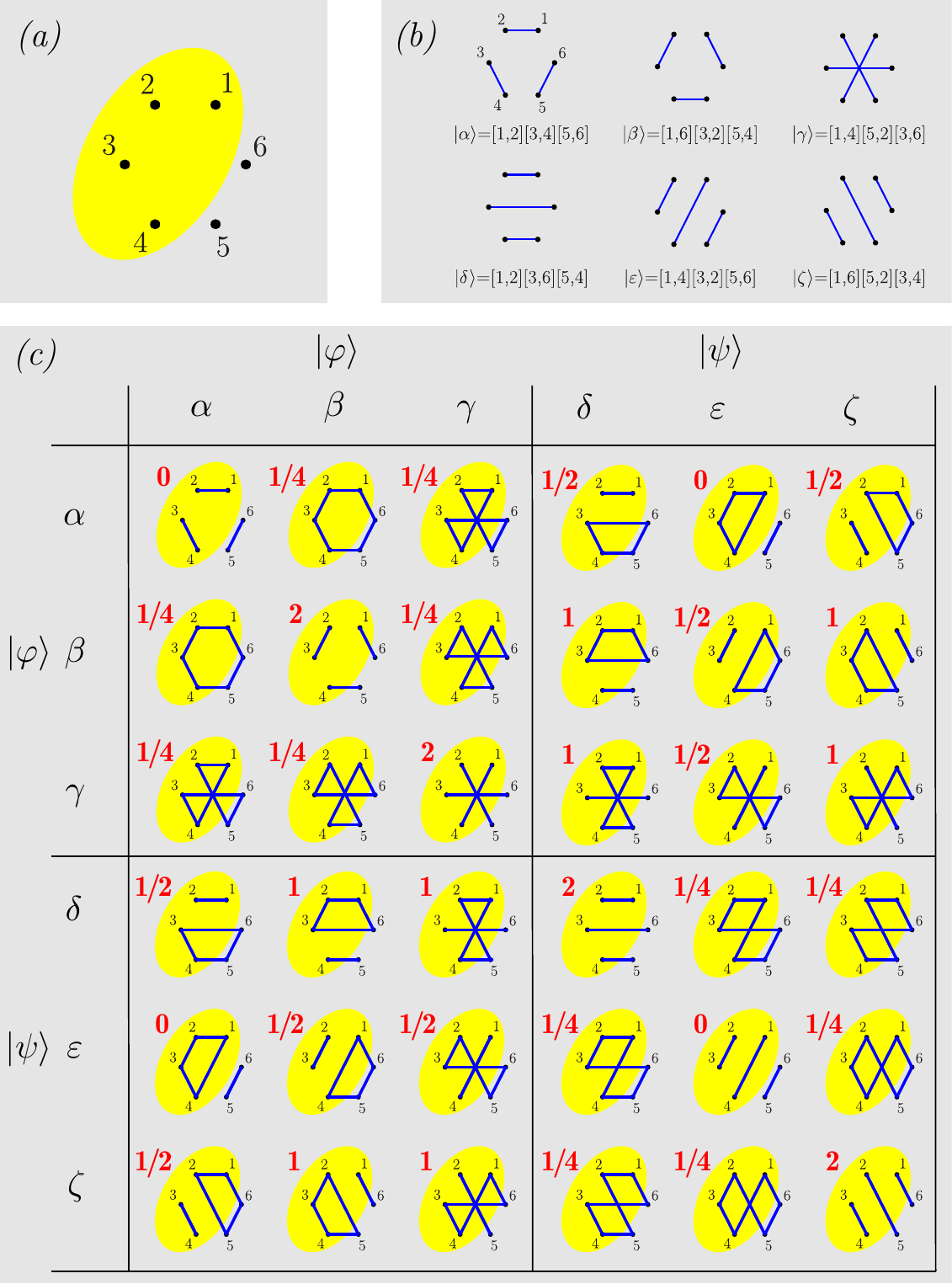,width=0.7\textwidth}}
\vspace*{8pt}
\caption{(Color online) 6-sites example showing that $S_{\text{loop}}$ introduced in Ref.~\protect\refcite{Sandvik2010} is not independent of the chosen decomposion of a given singlet state on the overcomplete bipartite VB basis. (a) Definition of the subsystems $A=(1,2,3,4)$ and $B=(5,6)$. (b) The six bipartite VB states. (c) Loop contributions $\langle v \vert \Lambda_{AB} \vert v' \rangle = ({\text{number of crossing loops}})\times\langle v \vert  v' \rangle$ (red numbers) involved in the computation of $S_{\text{loop}}$ for a single singlet state using either the representation $\vert \varphi \rangle = \vert \alpha \rangle + \vert \beta \rangle + \vert \gamma \rangle $ or $\vert \psi \rangle = \vert \delta \rangle + \vert \varepsilon \rangle + \vert \zeta \rangle $. Note that, due to overcompleteness,  $\vert \varphi \rangle = \vert \psi \rangle$.}
\label{fig:cexample}
\end{figure}

A simple argument based on a 6-sites example shows that $S_{\text{loop}}$ does not satisfy the criterion raised in the previous paragraph. Despite its manifest physical content\cite{Sandvik2010}, it is deeply tied to its stochastic definition inside the QMC scheme and cannot be consistently defined as a quantum average nor a projection. We consider two blocks $A=(1,2,3,4)$ and $B=(5,6)$ (see Fig.~\ref{fig:cexample}(a)) and the six bipartite states $\vert \alpha \rangle$, $\vert \beta \rangle$, $\vert \gamma \rangle$, $\vert \delta \rangle$, $\vert \varepsilon \rangle$ and $\vert \zeta \rangle$ (see Fig.~\ref{fig:cexample}(b)). Due to overcompleteness,  $\vert \varphi \rangle = \vert \alpha \rangle + \vert \beta \rangle + \vert \gamma \rangle $ and $\vert \psi \rangle = \vert \delta \rangle + \vert \varepsilon \rangle + \vert \zeta \rangle$ are two representations of the same singlet state, $\vert \varphi \rangle = \vert \psi \rangle$ . Note that the example is intentionally built such as the weights involded in the decompositions $\vert \varphi \rangle$ and  $\vert \psi \rangle$ are positive in order to stay closer to the situation encompassed by the projection QMC scheme. As shown in Fig.~\ref{fig:cexample}(c), using the notations of Ref.~\protect\refcite{Sandvik2010},  $\langle \varphi \vert \Lambda_{AB} \vert \varphi \rangle = 11/2$ while $\langle \varphi \vert \Lambda_{AB} \vert \psi \rangle = 6$, excluding the possibility for $S_{\text{loop}} = \langle  \Lambda_{AB} \rangle$ to be a quantum mechanical average.

Let us remark that even if this point raises a difficulty in the interpretation of $S_{\text{loop}}$ from a quantum mechanical point of view, this lessens in no way the numerical evidence that $S_{\text{loop}}$ and $S^{\text{vN}}$ probably have a very similar behavior.

\section{Numerical results}\label{num.part}

As an illustration, we now present some numerical results for $S^{\rm vN}$ and $S^{\rm VB}$ for frustrated spin-1/2 and spin-1 chains. 

\subsection{Heisenberg spin-1/2 chain}

Let us consider the frustrated spin-1/2 chain 
$\displaystyle
    H = \sum_{i=1}^L J_1 {\mathbf S}_i.{\mathbf S}_{i+1} + J_2 {\mathbf S}_i.{\mathbf S}_{i+2}
 $
where we set $J_1=1$ and vary $J_2$. The physics of this spin chain is well understood: for $J_2$ smaller than the critical value $J_2^c\simeq 0.241167$ (Ref.~\protect\refcite{Eggert}), the system displays antiferromagnetic quasi-long range order, with algebraically decaying spin correlations. For $J_2>J_2^c$, the system is located in a gapped dimerized phase which spontaneously breaks translation symmetry. In Ref.~\protect\refcite{Alet2010}, we have studied both $S^\mathrm{VB}$ and $S^\mathrm{vN}$  in both phases using large-scale DMRG technique.

To summarize our findings, we observe a similar behaviour for both EE in the gapless or gapped phases with respectively logarithmic increase or saturation. If we fit the prefactor of the logarithmic law in the gapless phase to a form $S^\mathrm{vN}=c_\mathrm{eff}/3 \ln(x') + K_1$ and $S^\mathrm{VB}=\gamma_\mathrm{eff} \ln(x') + K_2$, we obtain results very close to CFT predictions\cite{Calabrese04,Vidal,Jacobsen}: $c=1$ and $\gamma=4\ln(2)/\pi^2$. 
But even more interestingly is the fact that the values closest to these theoretical predictions are found to be \emph{precisely} at $J_2^c$, which indicates that both quantities can detect the quantum phase transition.

\subsection{Bilinear-biquadratic spin 1 chain}

How can we extend the definition of the VB occupation number (and VB entropy) to spin values $S>1/2$ ? Naively, one would try to count the number of 2-sites singlets formed by 2 spins $S$. While this is possible using the derivation outlined in Sec.~\ref{sec:VBdef}, it is crucial to remark that, in contrast to the spin $1/2$ case, such VBs do {\it not} form a basis of the singlet space of $N$ spins $S$: in fact they form a basis of SU(N) singlets\cite{Beach}, a symmetry which is only met by specific spin models with a symmetry higher than the usual SU(2).

Here we propose an alternative, more generic, definition which consists in computing the ``VB entropy of spins $1/2$'' in a spin $S$ system. Without loss of generality, we will describe this method for systems with spin $S=1$. At each lattice site $i$, decompose the spin $1$ ${\bf S}_i$ into $2$ spins 1/2 (denoted by ${\bf s}_{i,1}$ and ${\bf s}_{i,2}$), and consider the ensemble of VBs that can be formed by all these spins $1/2$. At each site, the Hilbert space is of size $4$ (instead of $3$ for a single spin $1$) and we will force the use of the physically relevant states by refusing that two spins $1/2$ form a singlet on the same lattice site. For systems on a bipartite lattice, this is easily done by taking the bipartite VB basis for the spins 1/2, where all ${\cal A}$ (${\cal B}$) spin 1 sites only host ${\cal A}$ (${\cal B}$) sites for the spins $1/2$. All spin $1/2$ VB states span the singlet sector of this fictitous spin $1/2$ system, and therefore the full singlet sector of the spin $1$ system. We can now count the occupation of these spin-$1/2$ VBs using the formula given above. It is easy to see that the ``spin $1/2$ VB occupation number'' $n^{S=1/2}_{i,j}$ of a spin for two spins $1$ at sites $i$ (in sublattice ${\cal A}$) and $j$ (in sublattice ${\cal B}$) in a wavefunction $\vert \Psi \rangle$, is obtained by putting together all spin-1/2 ${\bf s}$ operators : $s^+_{i,1}s^-_{j,1}+s^+_{i,1}s^-_{j,2}+s^+_{i,2}s^-_{j,1}+s^+_{i,2}s^-_{j,2}$, resulting in 
\begin{equation}
n^{S=1/2}_{i,j} (\vert \Psi \rangle)=  - \frac{\langle R_1 \vert \hat{S}^{+}_i \hat{S}^{-}_j \vert \Psi \rangle}{\langle R_1 \vert \Psi \rangle}.
\end{equation}
Here $\vert R_1\rangle$ is the reference state for $S=1$ defined in Sec.~\ref{sec:VBdef}. Note the resemblance with the counting of spin-$1$ VB, except for the factor $2$. Indeed, there is at most one spin-$1$ VB between two spin $1$ sites, whereas there can be up to two spin-$1/2$ VBs. With this definition, we can now compute the ``VB entropy of spins $1/2$''  by counting how many spins $1/2$ VBs emerge from a given block size of a spin $1$ system, multiplied by $\ln(2)$. It is important to remark that the spin $1/2$ artificial system is useful only to identify our computations, but that all the simulations and numerical evaluations can be made directly in the spin-$1$ model and formalism.

For illustration, we consider the spin-1 chain with bilinear and biquadratic terms
\begin{equation}
{\cal H} = \sum_{i} \cos \theta \, {\mathbf S}_i.{\mathbf S}_{i+1} + \sin \theta \, ({\mathbf S}_i.{\mathbf S}_{i+1})^2,
\end{equation}
which phase diagram is well known thanks to  analytical and numerical studies.\cite{spin1}

Here, we focus on one side of the gapped Haldane phase, for instance by choosing $\theta=0$ (Heisenberg point), $\theta=10^\circ$ and $\theta=\arctan(1/3)$ which corresponds to the famous AKLT point with an exact ground-state\cite{AKLT} and a short correlation length $\xi =1/\ln 3$. On the other side, we want to see whether $S^\mathrm{VB}$  can detect the phase transition that occurs at $\theta=-\pi/4$, the Babujian-Takhtajan (BT) point, where criticality is known to be described by a $SU(2)_2$ CFT with central charge $c=3/2$.\cite{BT}
In what follows we use DMRG to compute the ground-state of ${\cal H}$ and the von Neumann and spin-$1/2$ VB entropies for chains of size $L=64$ with PBC.


\begin{figure}[htbp]
\centerline{\psfig{file=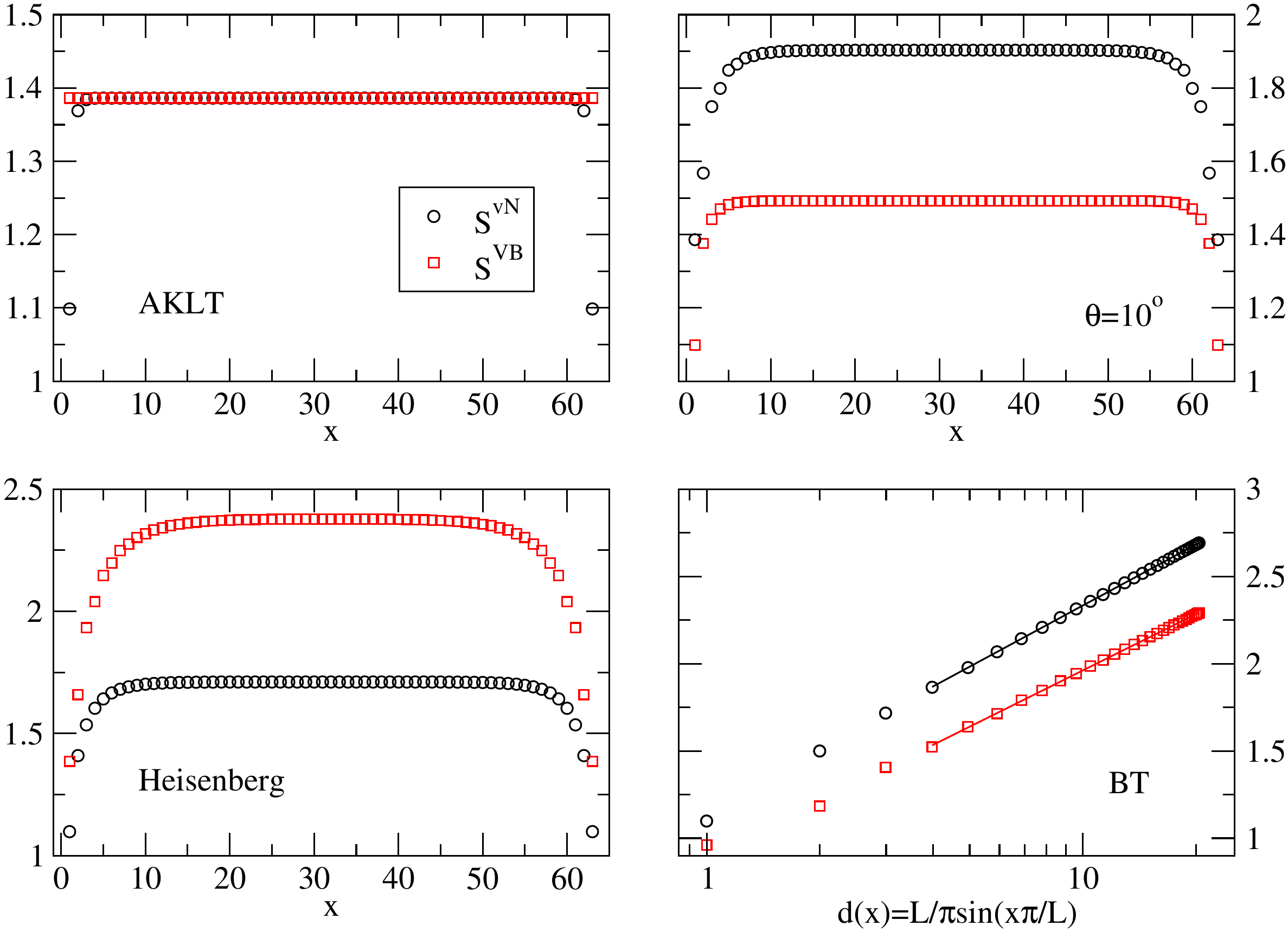,width=\textwidth}}
\vspace*{8pt}
\caption{Scaling as a function of block size $x$ of the von Neumann (black circles) and spin-$1/2$ VB (red squares) entropies for the  bilinear-biquadratic spin $1$ chain for various values of $\theta$ : $\theta=\arctan(1/3)$ (AKLT point), $\theta=10^\circ$, $\theta=0$ (Heisenberg point), and $\theta=-\pi/4$ (BT point). In the last case, entropies are plotted versus conformal distance $d(x)=L/\pi \sin (x\pi/L)$ on a semi-log scale and lines are best fits to a logarithmic divergence (see text). We used $L=64$ and PBC.}
\label{fig:DMRG}
\end{figure}

At the AKLT point, $S^\mathrm{vN}$ saturates to $2\ln 2$ rapidly (with the same correlation length $\xi$ as the bulk one) as was known from previous study.\cite{Geraedts}
The VB entropy of spins $1/2$ is constant for all block sizes at the AKLT point, and the constant is also exactly equal to $2 \ln (2)$: this is easily understood noting the VB solid structure in the AKLT ground-state where there is exactly one spin $1/2$ VB on every lattice link. For the Heisenberg point or $\theta=10^\circ$, saturations of both entropies can be readily seen too, but now there is no direct relation between the constants towards wich both $S$ and $S^\mathrm{VB}$ saturate.

For the BT point, $S^\mathrm{vN}$ is expected to show logarithmic behaviour with a slope $c/3$ (since we are using PBC), with $c=3/2$ from the known criticality. Indeed, we observe a very good agreement with our data. 
The VB entropy of spins $1/2$ also exhibits a logarithmic divergence with block size, with a slighlty different prefactor $\gamma_{SU(2)_2}$, exactly as in the spin-1/2 $SU(2)$ case. Theoretical work by Saleur and Jacobsen\cite{Saleur}, made along the lines of Ref.~\protect\refcite{Jacobsen}, predict $\gamma_{SU(2)_2}=2.\gamma=8\ln(2)/\pi^2\sim 0.562$, again close but different from $c/3=0.5$. Our best fit gives $\gamma_{SU(2)_2}=0.667$, slightly different from the theoretical value. This discrepancy is likely due to logarithmic corrections due to the enhanced symmetry at this point, as was found in the spin $1/2$ Heisenberg case\cite{Alet2010}.

\section{Conclusion}
It is now obvious that entanglement measurements can shed light on the nature of the physical state, without having any a priori input. While the computation of  $S^\mathrm{vN}$ for large interacting 2d systems remain a very difficult task, there has been several recent progress using large-scale QMC simulations of spin systems in the VB basis. We have reviewed some proposals and discussed especially if these new estimators are valid quantum observables or projections. 
We focused on the VB EE, which can be either defined as a geometrical quantity, or as a projection onto a particular state, and which has been shown to exhibit several interesting features for gapped or gapless spin-1/2 phases. We have described how it can be extended to arbitrary spin and provide numerical data for frustrated spin-1 chain, where a critical state can be detected using this tool. Clearly, more work should be devoted to understand similarities and differences between these various estimators. 

\section*{Acknowledgments}

We thank K. Beach, N. Laflorencie, and I. McCulloch for their participation in related work. We thank I. McCulloch for providing the DMRG code used in generating the results of Fig.~\ref{fig:DMRG}, as well as H. Saleur and J. Jacobsen for communicating their unpublished results\cite{Saleur}. We are also grateful to A. Sandvik for stimulating discussions. We thank CALMIP for allocation of CPU time. 
This work is supported by the French ANR program ANR-08-JCJC-0056-01. SC also thanks IUF for funding.


\end{document}